\documentstyle[12pt]{article}

\advance\voffset by -1.5cm
\advance\hoffset by -1.8cm
\input amssym.def
\input amssym.tex
\def\be{\begin{equation}}
\def\ee{\end{equation}}

\def\Zop{{\Bbb Z}}

\def\pmb#1{\setbox0=\hbox{#1}%
 \kern-.025em\copy0\kern-\wd0
 \kern.05em\copy0\kern-\wd0
 \kern-.025em\raise.0433em\box0 }

\def\O{{\cal O}}

\def\A{{\cal A}}

\def\H{{\cal H}}

\def\F{{\cal F}}

\def\ct{{\tilde{c}}}
\def\tG{{\widetilde G}}

\def\tS{{\tilde S}}
\def\tG{{\tilde G}}

\def\3{\ss}
\def\sq{\hbox{\rlap{$\sqcap$}$\sqcup$}}
\def\qed{\ifmmode\sq\else{\unskip\nobreak\hfil
\penalty50\hskip1em\null\nobreak\hfil\sq
\parfillskip=0pt\finalhyphendemerits=0\endgraf}\fi}
\def\half {\frac{1}{2}}
\def\thalf {\frac{3}{2}}
\def\fhalf {\frac{5}{2}}

\def\bbbz {{\sf Z\!\!Z}}

\def\Nu{{{\cal N}}}
\def\bbbone {{\mathchoice {{\rm 1\mskip-4mu l}} {{\rm 1\mskip-4mu l}}
{{\rm 1\mskip-4.5mu l}} {{\rm 1\mskip-5mu l}}}}
\def\bbbc{{\mathchoice {\setbox0=\hbox{$\displaystyle\rm C$}\hbox{\hbox
to0pt{\kern0.4\wd0\vrule height0.9\ht0\hss}\box0}}
{\setbox0=\hbox{$\textstyle\rm C$}\hbox{\hbox
to0pt{\kern0.4\wd0\vrule height0.9\ht0\hss}\box0}}
{\setbox0=\hbox{$\scriptstyle\rm C$}\hbox{\hbox
to0pt{\kern0.4\wd0\vrule height0.9\ht0\hss}\box0}}
{\setbox0=\hbox{$\scriptscriptstyle\rm C$}\hbox{\hbox
to0pt{\kern0.4\wd0\vrule height0.9\ht0\hss}\box0}}}}

\begin{document}
\thispagestyle{empty}
\def\thefootnote{\fnsymbol{footnote}}
\begin{flushright}
DAMTP-96-65 \\
hep-th/9607036
\end{flushright}
\vspace{2.0cm}

\begin{center}

{\Large {\bf Fusion of twisted representations}}
\vspace{2.0cm}

{\large Matthias R. Gaberdiel} 
\footnote{e-mail: M.R.Gaberdiel@damtp.cam.ac.uk}
\footnote{Address from 1 Sep 1996: Department of Physics,
Harvard University, Cambridge MA, 02138, USA.} \\ 
{Department of Applied Mathematics and Theoretical Physics\\ 
University of Cambridge, Silver Street 
\\ Cambridge, CB3 9EW, U.\ K.\ }
\vspace{0.5cm}

July 1996
\vspace{3.0cm}

{\bf Abstract}
\end{center}

{\leftskip=2.4truecm
\rightskip=2.4truecm

The comultiplication formula for fusion products of untwisted
representations of the chiral algebra is generalised to include
arbitrary twisted representations. We show that the formulae 
define a tensor product with suitable properties, and determine the
analogue of Zhu's algebra for arbitrary twisted representations. 

As an example we study the fusion of representations of the Ramond
sector of the $N=1$ and $N=2$ superconformal algebra. In the latter
case, certain subtleties arise which we describe in detail.

}

\newpage

\setcounter{footnote}{0}
\def\thefootnote{\arabic{footnote}}

\setcounter{page}{1}
\section{Introduction}
\renewcommand{\theequation}{1.\arabic{equation}}
\setcounter{equation}{0}

Some years ago, Richard Borcherds proposed that fusion in (chiral)
conformal field theory should be regarded as the ring-like tensor
product of representations of a {\it quantum ring}, a generalisation
of rings and vertex algebras, for which the holomorphic fields of a
chiral theory form the natural example. From this point of view,
conformal field theory is just the representation theory of this
quantum ring, and fusion corresponds to taking the canonical ring-like
tensor product of modules. A somewhat similar definition has now also
been given by Huang and Lepowsky \cite{HL}.

In \cite{MG1}, this approach was formulated for untwisted
representations in terms of the chiral algebra, the mode expansion of
the holomorphic fields. Using standard arguments of conformal field
theory, two different actions (comultiplications) of the chiral
algebra on the vector space tensor product of representations were
derived. The fusion product was then defined as the (ring-like)
quotient of the original tensor product by all relations which enforce
the equality of the two different actions.

The two different actions were given explicitly, and as a consequence
it was possible to prove various properties of the resulting fusion 
product. It was also checked for a number of example \cite{MG1,MG2}
that the resulting fusion rules reproduce the know restrictions. 

It has become apparent that this approach to fusion does not only give
a satisfactory description from a conceptual point of view, but that it
also allows one to prove various structural properties of fusion
\cite{Nahm}. In addition, it provides a rather powerful method for the
calculation of the fusion rules. This seems, in particular, to be the
case in the situation, where the correlation functions contain
logarithms \cite{Gurarie}, whence the fusion product is not completely
reducible \cite{Nahm,GK1,GK2}. 
\smallskip

The above analysis was only performed for the situation in which all
representations are untwisted. For these representations, the moding
of the chiral algebra is the same as in the vacuum representation, and
the interpretation of the fusion product as a tensor product is
rather suggestive. In this paper we show that this interpretation is not
limited to this simple situation, and that the general case of twisted
representations can be treated similarly. We believe that this is
important from a conceptual point of view, but, as indicated before,
it should also allow us to calculate fusion rules fairly efficiently. 
As an example, and in order to check our formulae, we derive the
fusion rules of the $N=1$ superconformal algebra in this way.

Part of the original motivation was to understand the $N=2$ fusion
rules for the Ramond and Neveu-Schwarz sectors, and consequently,
quite a substantial part of the paper deals with this problem.
One of the central themes is to describe the way in which the fusion
rules of the $N=2$ algebra respect the automorphism symmetry from the
spectral flow \cite{SchS}. A number of subtleties arise in this
context, and we explain them in detail. Some of them seem to have been
overlooked so far.
\smallskip

The paper is organised as follows. In the next section we introduce
the relevant notation, and derive the generalised comultiplication
formula. We also indicate how some of its properties can be shown. In
section 3, we use the explicit formulae to find the analogue of Zhu's
algebra \cite{Zhu} for the twisted case, thereby generalising further
the analysis of \cite{KacWang} and \cite{DLM}. We also explain how to
define the special subspace \cite{Nahm} in the more general
situation. In the reminder of the paper we calculate explicitly the
fusion rules for the Ramond and Neveu-Schwarz sector of the $N=1$
(section 4) and $N=2$ (section 5) superconformal algebra.

\section{The general comultiplication formula}
\renewcommand{\theequation}{2.\arabic{equation}}
\setcounter{equation}{0}

\subsection{Definitions and Notation}
\renewcommand{\theequation}{2.1.\arabic{equation}}

Let $S^i(w), i=1, \ldots,n$ denote (a suitable subset of) the
holomorphic quasiprimary fields of a conformal field theory, where
$S^i$ has conformal weight $h_i\in\Zop/2$. We define the vacuum sector
$\H_0$ of the corresponding chiral conformal field theory as a certain
completion of the vector space spanned by products of the form
\be
\label{dense}
S^{j_1} (w_1) \cdots S^{j_n} (w_n) \; \Omega \,, 
\ee 
where $\Omega$ is the vacuum vector, and $w_i\neq w_j$ for $i\neq j$
(see \cite{MG3} and \cite{GG} for more details). We can associate to
a state $\Psi\in\H_0$ a field $V(\Psi,z)$ which then satisfies
$V(\Psi,0) \Omega = \Psi$.

The correlation functions of the holomorphic fields are meromorphic
(single-valued) functions. This implies, in particular, that the
holomorphic fields satisfy an operator product expansion  
\be
\label{ope}
S^1(w) S^2(z) = \sum_{l\in \Zop + h_1} (w-z)^{l-h_1} \;
V(S^1_{-l} S^2(0) \Omega, z) \,,
\ee
where $|w - z|$ is sufficiently small. If the spectrum of the scaling
operator $L_0$ is positive (as we shall assume), then the sum in
(\ref{ope}) is bounded by $l\geq -h_2$.  

Because of (\ref{ope}) it is possible to expand the holomorphic
fields in terms of modes as ({\it cf.}  \cite{Peter89}) 
\be
\label{mode}
S(w)=\sum_{l\in\Zop + h} w^{l-h} \; S_{-l}\,.
\ee
The singular part of the operator product expansion gives rise to
commutation relations for the modes; these generate an infinite 
dimensional algebra (typically a $W$-algebra) which is called the 
{\it chiral algebra} $\A$ of the conformal field theory.

The other sectors of the chiral theory can be interpreted as
representations $\H$ of the vacuum sector $\H_0$. A representation is
defined by a set of amplitudes, involving an arbitrary number of
holomorphic and two special fields, which we may imagine to correspond
to a state in a representation $\H$, and one in its dual $\H^*$. The
crucial property which has to be satisfied by these amplitudes is that
they respect the relations coming from the vacuum amplitudes, and in
particular (\ref{ope}) (see again \cite{MG3} and \cite{GG} for more
details). Furthermore, the amplitudes have to possess suitable
analytic properties, and the short distance behaviour of the
holomorphic field $S^i(w)$ with each of the two special fields has to
be of the form 
\be
\label{action}
S^i(w) V(\psi,z) = \sum_{l\geq -h_{\psi}} (w-z)^{l-h_i} V(\psi^i_l, z) \,,
\ee
where $\psi^i_l$ defines another set of allowed amplitudes,
$l\in\Zop+h_i+\alpha^i$, and $h_{\psi}$ is some finite number which is
independent of $S^i$. We can use this to define an action of the modes of
$S^i$ on states in $\H$, by defining (in analogy with (\ref{ope}))
$S^i_{-l}\;\psi:=\psi^i_l$. As the correlation functions satisfy the
conditions coming from (\ref{ope}), this action defines a
representation of the chiral algebra. Furthermore, the null-relations
of the vacuum sector are respected, and the resulting representation
is of positive energy, as the sum is bounded from below.

There are two different types of representations, usually referred to
as {\it untwisted} and {\it twisted}. In the untwisted case, the sum
in (\ref{action}) runs over $l\in\Zop + h_i$ (for all $S^i$), and the
correlation functions are meromorphic functions of the arguments of
the holomorphic fields. (In particular, the modes of the holomorphic
fields have then the same moding as in the vacuum sector.) In the
twisted  case, there exists at least one holomorphic field $S^j$ which
is not single-valued, and for which the sum runs over
$l\in\Zop+h_j+\alpha^j$, where $\alpha^j\not\in\Zop$. The 
monodromy of the field $S^j$ around states in $\H$ is characterised 
by $\alpha^j \; (\mbox{mod}\; \Zop)$. We shall call $\alpha^j$ in
the following the {\it twist} of the representation (with respect to
the field $S^j$).
\smallskip

An important aspect of the chiral theory is to understand the
correlation functions involving more than two non-holomorphic
fields. Because of the analyticity of the amplitudes it is sufficient to
analyse the situation where there are three such fields, as the
general case can be reduced inductively to this case. These
three-point functions are largely determined by the constraints which
come from the compatibility with the chiral algebra. Indeed, in
\cite{MG1} the point of view was put forward, that one should think of
the three-point functions as the decomposition of a ring-like tensor
product of two of the representations into the conjugate of the
third. This was developed further in \cite{MG2}, where the general
formula for the action of the chiral algebra on the tensor product was
derived for the case that both representations are untwisted. In this
paper, we want to find the generalisation of this formula to the case
where the representations in question are not necessarily
untwisted. We shall see that the same interpretation can be given to
these products. This demonstrates that this approach is rather
general.

\subsection{Derivation of the Comultiplication Formula}
\renewcommand{\theequation}{2.2.\arabic{equation}}

Let $\psi_1$ and $\psi_2$ be two vectors in representations $\H_1$ and
$\H_2$, respectively, and let the twist of the representation $\H_i$
(with respect to the field $S$) be described by $\alpha_i$. The
product of the two fields will then define a representation
whose twist (with respect to $S$) is given by
$\alpha=\alpha_1+\alpha_2$. We want to derive a formula for the action
of modes of $S$ (in a representation with twist $\alpha$) on the product
of the two states. This means that we want to calculate the contour
integral  
\be
\label{int}
\oint_{C} dw \; w^{l+h+\alpha-1} \;S(w) \;V(\psi_1,z_1)
V(\psi_2,z_2)\;\Omega \,, 
\ee
where $C$ is a contour encircling the two insertion points. This will
give a formula for the comultiplication
$\Delta_{z_1,z_2}(S_l)(\psi_1\otimes\psi_2)$. 

Unfortunately, (\ref{int}) only converges in correlation functions
with vectors of finite energy (at infinity), and thus it is not
possible to evaluate the integral independently of the state at
infinity. To circumvent this problem we consider, following Friedan
\cite{Friedan}, a slightly modified integral  
\be
\label{intmod}
\Delta_{z_1,z_2}(\tS_l)(\psi_1\otimes\psi_2) = 
\oint_{C} dw \; w^{l+h+\alpha-1} (w-z_1)^{-\alpha_1} 
(w-z_2)^{-\alpha_2} \;S(w) \;V(\psi_1,z_1) V(\psi_2,z_2)\;\Omega
\ee
for which this problem does not arise. It is possible to express $S_l$
in terms of $\tS_m$ by expanding the function $(w-z_1)^{-\alpha_1}
(w-z_2)^{-\alpha_2}$ for large $w$, 
\be
\label{tilde}
\tS_l = \sum_{p_1=0}^{\infty} \sum_{p_2=0}^{\infty}
\left( \begin{array}{c} -\alpha_1 \\ p_1 \end{array} \right)
\left( \begin{array}{c} -\alpha_2 \\ p_2 \end{array} \right)
(-z_1)^{p_1} (-z_2)^{p_2}  S_{l - (p_1 + p_2)} \,,
\ee
and similarly to write $S_l$ in terms of $\tS_m$. It is therefore
equivalent to give an action of the chiral algebra in 
terms of the modes $\tS_m$ or in terms of the modes $S_l$.

The important property of the relation between $S_l$ and $\tS_m$
is that it is of the form   
\be
\label{property}
\tS_l = S_l + \sum_{m<l} a_m S_m \hspace*{1cm}
S_l = \tS_l + \sum_{m<l} b_m \tS_m \,,
\ee
which will be important in later calculations. This depends on the
fact that $\alpha=\alpha_1 + \alpha_2$, and we shall always choose
$\alpha$ in this way.  
\smallskip

To find an explicit formula for $\Delta_{z_1,z_2}(\tS_l)$, let us
(as in \cite{MG1}) introduce the function
\be
\F(w; z_1,z_2) := (w-z_1)^{-\alpha_1} (w-z_2)^{-\alpha_2}
\langle \phi,  S(w) V(\psi_1,z_1) V(\psi_2,z_2) \rangle \,,
\ee
where $\phi$ is any vector in the dense subspace of finite energy
vectors of the chiral conformal field theory. (It is clear that $\F$
can only be different from zero, if $\phi$ is in a representation with
twist $\alpha$.) By construction, this function (as a function of $w$)
has no branch cuts, and its poles are given by 
\be
\label{poles}
\sum_{l\leq \alpha_1+h-1} (w-z_1)^{l-h-\alpha_1} (w-z_2)^{-\alpha_2}
\langle (S_{-l} \otimes \bbbone) \rangle
+ \varepsilon_1 
\sum_{m\leq \alpha_2+h-1} (w-z_1)^{-\alpha_1} (w-z_2)^{m-h-\alpha_2} 
\langle (\bbbone\otimes S_{-m}) \rangle \,,
\ee
where we use the short-hand notation
\be
\langle (S_{-l} \otimes \bbbone) \rangle = 
\langle \phi, V(S_{-l}\psi_1,z_1) V(\psi_2,z_2) \rangle \,,
\ee
and likewise for $\langle (\bbbone\otimes S_{-m}) \rangle$. Here
$\varepsilon_1$ is a complex number which is defined by the identity
\be
S(w) V(\psi_1,z_1) = \varepsilon_1 V(\psi_1,z_1) S(w) \,,
\ee
where $|w| > |z_1|$, and the right-hand-side is to be understood as the
(clockwise) analytic continuation from $|w| < |z_1|$ ({\it cf.}
\cite{Peter89}). In particular, we have
\be
\varepsilon_i = e^{\pm \pi i \alpha_i} \,.
\ee

As in \cite{MG1}, we can then define the regular part of $\F$ by
subtracting the poles (\ref{poles}) from an expansion of $\F$, using
the short distance expansion of $S$ with $V(\psi_2,z_2)$
\be
\label{expansion}
\F(w; z_1,z_2) =  \varepsilon_1 
\sum_{m\in \Zop+h+\alpha_2} (w-z_1)^{-\alpha_1} (w-z_2)^{m-h-\alpha_2} 
\langle (\bbbone\otimes S_{-m}) \rangle \,.
\ee
We obtain 
\pagebreak

$$
\F_{reg}(w; z_1,z_2) = 
- \sum_{l\leq \alpha_1+h-1} (w-z_1)^{l-h-\alpha_1} (w-z_2)^{-\alpha_2}
\langle (S_{-l} \otimes \bbbone) \rangle \hspace*{4cm} 
$$
$$ \hspace*{3cm} 
+ \varepsilon_1 
\sum_{m\geq \alpha_2+h} (w-z_1)^{-\alpha_1} (w-z_2)^{m-h-\alpha_2} 
\langle (\bbbone\otimes S_{-m}) \rangle \,.
$$
It is clear that the regular part $\F_{reg}$ of $\F$ only contributes
to (\ref{intmod}) for  $n+h+\alpha \geq 0$. For $n+h+\alpha \geq 1$,
therefore only the singular part (\ref{poles}) of $\F$ contributes to
the comultiplication formula (\ref{intmod}), and we easily find
(for $n+h+\alpha \geq 1$)
\begin{eqnarray}
{\displaystyle \Delta_{z_1,z_2}(\tS_{n})} & = &
{\displaystyle \sum_{l=1-h-\alpha_1}^{\infty} \sum_{q=0}^{\infty}
\left( \begin{array}{c} n+h+\alpha-1 \\ q \end{array} \right)
z_1^{n+h+\alpha-1-q}} \nonumber \\
& & \hspace*{0.4cm} {\displaystyle \left( \begin{array}{c} -\alpha_2 \\
h+\alpha_1+l-1-q \end{array} \right)
(z_1 - z_2)^{-\alpha_2 - h - \alpha_1 - l + 1 + q}
\left(S_{l} \otimes \bbbone\right)  }
\nonumber \\
\label{chir1}
& &  {\displaystyle +\, \varepsilon_{1}
\sum_{m=1-h-\alpha_2}^{\infty} \sum_{q=0}^{\infty}
\left( \begin{array}{c} n+h+\alpha-1 \\ q \end{array} \right)
z_2^{n+h+\alpha-1-q}} \nonumber \\
& & \hspace*{0.4cm} {\displaystyle \left( \begin{array}{c} -\alpha_1 \\
h+\alpha_2+m-1-q \end{array} \right) 
(z_2 - z_1)^{-\alpha_1 - h -\alpha_2 - m + 1 + q}  
\left(\bbbone\otimes S_{m}\right)\,. }
\end{eqnarray}

For $n+h+\alpha \geq 0$, both the regular and the singular part
contribute, and the explicit formula therefore depends on the expansion
chosen in (\ref{expansion}). As in \cite{MG1}, there will be two
different formulae, whose action on states has to agree in all
correlation functions. Using the expansion with $V(\psi_2,z_2)$ (as in
(\ref{expansion})), we get (for $n-h-\alpha\geq 0$)
\begin{eqnarray}
{\displaystyle \Delta_{z_1,z_2}(\tS_{-n})} & = &
{\displaystyle \sum_{l=1-h-\alpha_1}^{\infty} \sum_{q=0}^{\infty}
\left( \begin{array}{c} n+l-\alpha+\alpha_1-1-q \\ n-h-\alpha \end{array} \right)
(-1)^{h+\alpha_1+l-1-q} z_1^{l-n+\alpha-\alpha_1+q}} \nonumber \\
& & \hspace*{2.4cm} {\displaystyle \left( \begin{array}{c} -\alpha_2 \\
q \end{array} \right)
(z_1 - z_2)^{-\alpha_2 - q}
\left(S_{l} \otimes \bbbone\right)  }
\nonumber \\
\label{chir21}
& &  {\displaystyle +\, \varepsilon_{1}
\sum_{m=1-h-\alpha_2}^{\infty} \sum_{q=0}^{\infty}
\left( \begin{array}{c} h+\alpha_2+m-1+q \\ q \end{array} \right)
z_2^{q}} \nonumber \\
& & \hspace*{2.4cm} {\displaystyle \left( \begin{array}{c} -\alpha_1 \\
n+m+\alpha_2-\alpha+q \end{array} \right) 
(- z_1)^{-m-n-q}  
\left(\bbbone\otimes S_{m}\right)  } \nonumber \\
& &  {\displaystyle +\, \varepsilon_{1}
\sum_{m=\alpha_2+h}^{\infty} \sum_{q=0}^{n-h-\alpha}
\left( \begin{array}{c} m-h-\alpha_2 \\ q \end{array} \right)
(-z_2)^{m-h-\alpha_2-q}} \nonumber \\
& & \hspace*{2.4cm} {\displaystyle \left( \begin{array}{c} -\alpha_1 \\
n-h-\alpha-q \end{array} \right) 
(- z_1)^{-n+h+\alpha-\alpha_1 + q}  
\left(\bbbone\otimes S_{-m}\right)\,.  }
\end{eqnarray}
Here we have chosen to expand the functions as power series of $z_2$
about $0$, so that the limit $z_2\rightarrow 0$ is 
well-defined. It is then easy to see that they reduce to the
expressions given in \cite{MG2} for
$\alpha_1=\alpha_2=\alpha=0$.

As mentioned before, there exists a second formula which can be
obtained using the expansion of $S$ with $V(\psi_1,z_1)$ in
(\ref{expansion}). Denoting, as in \cite{MG1} the relevant formula by 
$\widetilde{\Delta}_{z_1,z_2}(\tS_{-n})$, it is given by
(\ref{chir21}) upon exchanging $z_1 \leftrightarrow z_2$,  $\alpha_1
\leftrightarrow \alpha_2$, and the two factors in the tensor
product\footnote{The formula given in \cite{MG2} has the wrong
$\varepsilon$ factors. This is corrected in \cite{EG}.}.
The fusion product is then defined as the quotient of the vector space
tensor product by the relations which come from the equality of 
$\Delta_{z_1,z_2}(\tS_{-n})$ and
$\widetilde{\Delta}_{z_1,z_2}(\tS_{-n})$. 
\smallskip

These explicit formulae are useful for actual calculations as
we shall demonstrate in the following sections. For more structural
considerations, it is sometimes better to use a formula, where the
various residues have not yet been evaluated
\begin{eqnarray}
\label{residue}
{\displaystyle \Delta_{z_1,z_2}(\tS_{-n})} & = &
{\displaystyle
\sum_{l} Res_{w=z_1} \left( (w-z_2)^{-\alpha_2} w^{n+h+\alpha-1}
(w-z_1)^{-(l+h+\alpha_1)}\right) (S_l \otimes \bbbone)} \nonumber \\
& = &
{\displaystyle
\sum_{m} Res_{w=z_2} \left( (w-z_1)^{-\alpha_1} w^{n+h+\alpha-1}
(w-z_2)^{-(l+h+\alpha_2)}\right) (\bbbone \otimes S_m)\,.\hspace*{0.5cm}}
\end{eqnarray}
Here the residue means
\be
Res_{w=z_1} = \oint_{C_1} dw \hspace*{2cm} 
Res_{w=z_2}=\oint_{C_2} dw \,, 
\ee
where the contour $C_1$ encircles $z_1$, but not $z_2$ or $0$, and 
the contour $C_2$ encircles $z_2$ and $0$, but not $z_1$.

\subsection{Some Properties}
\renewcommand{\theequation}{2.3.\arabic{equation}}

As in \cite{MG1,MG2}, we expect that the comultiplication formula will
have a natural transformation property under translation. This will
indeed turn out to be true, but the situation here is slightly more
complicated, as the $\tS_l$ (with respect to which we have formulated
our formulae) also depend on the two insertion points (as is 
obvious from (\ref{tilde})). To be more explicit about this dependence
let us denote for the moment the modes in (\ref{tilde}) by 
$\tS_l^{z_1,z_2}$. Then it is easy to see that in the sector described 
by the twist $\alpha$ we have the transformation law
\begin{equation}
\label{trans}
e^{u L_{-1}}\; \tS_{l}^{z_1,z_2} \; e^{-u L_{-1}} = \left\{
\begin{array}{ll}
{\displaystyle \sum_{m=1-\alpha-h}^{l}
\left( \begin{array}{c} l+h+\alpha-1 \\ m+h+\alpha-1 \end{array}
\right) (-u)^{l-m} \tS_{m}^{z_1+u,z_2+u}} & \mbox{if $l\geq 1-h-\alpha$} \\
{\displaystyle \sum_{m=-l}^{\infty}
\left( \begin{array}{c} m-h-\alpha \\ -m-h-\alpha \end{array} \right)
u^{m+l} \tS_{-m}^{z_1+u,z_2+u}} & \mbox{if $l\leq -h-\alpha$}\,.
\end{array}
\right.
\end{equation}
This is respected by the comultiplication, {\it i.e.} we have
\be
\label{cotrans}
\Delta_{z_1+u,z_2+u} 
\left(e^{u L_{-1}}\; \tS_{l}^{z_1,z_2} \; e^{-u L_{-1}}\right)
= \Delta_{z_1,z_2} \left(\tS_{l}^{z_1,z_2} \right) \,,
\ee
and likewise for the other comultiplication. This implies in
particular that the formulae can at most depend on the difference
of the two insertion points.
\medskip

Next, we want to discuss the coassociativity of the fusion product. In
order to analyse this it is advantageous to use the form of the 
comultiplication formula in which the various residues have not yet
been explicitly evaluated (\ref{residue}). We also have to be careful
to convert the $S_l$ modes into the appropriate $\tS_m$
modes. Keeping this in mind it is then not hard to see that 
\begin{equation}
\label{coass}
\left( \Delta_{\zeta_{2} - w,\zeta_{1}-w}\otimes\bbbone \right)
\circ \Delta_{w,z} =
\left( \bbbone\otimes \Delta_{\zeta_{1}-w,z-w} \right) \circ
\Delta_{\zeta_{2},w},
\end{equation}
and similarly for $\widetilde{\Delta}$ (see \cite{MG1}). This
establishes then that the fusion product is coassociative up to 
equivalence. 

Finally, we should mention that by the same arguments as in
\cite{MG1,MG2}, the comultiplication formulae must satisfy
\be
{}[ \Delta_{z_1,z_2} (a),  \Delta_{z_1,z_2} (b)] = 
\Delta_{z_1,z_2} ([a,b])
\ee
for all $a,b\in\A$. However, as the formulae are formulated in terms
of $\tS_m$ (which do not satisfy simple commutation relations), this
is rather difficult to check explicitly.

\section{Zhu's algebra}
\renewcommand{\theequation}{3.\arabic{equation}}
\setcounter{equation}{0}

\subsection{Derivation of Zhu's algebra}
\renewcommand{\theequation}{3.1.\arabic{equation}}

It was pointed out in \cite{EG} that it is possible to rederive Zhu's
algebra \cite{Zhu} using an approach based on the comultiplication
formulae. In this section we shall follow this idea to find the
generalisation of Zhu's algebra for the general twisted case. Zhu's
algebra has been generalised to the fermionic (untwisted) case in
\cite{KacWang}, and to the twisted bosonic case in \cite{DLM}.  

The main idea of the derivation is to consider the fusion product of a
representation ${\cal H}$ (which we assume without loss of generality
to be inserted at $z_2=0$) with the vacuum representation at
$z_1=z$. Then we have $\alpha_1=0$, $z_2=0$, and writing $z=z_1$,
$\alpha=\alpha_2$ the formulae simplify considerably 
\be
\label{chir10}
\Delta_{z,0}(\tS_{n}) = \sum_{l=1-h}^{\infty} 
\left( \begin{array}{c} n+h-1 \\ l+h-1 \end{array} \right)
z^{n-l} \left(S_{l} \otimes \bbbone\right)  +\, 
\varepsilon_{1} \left(\bbbone\otimes S_n\right)\,,
\ee
where $n\geq 1-h-\alpha$, and
\be
\label{chir20}
\Delta_{z,0}(\tS_{-n})  = 
\sum_{l=1-h}^{\infty} 
\left( \begin{array}{c} n+l-1 \\ l+h-1 \end{array} \right)
(-1)^{l+h-1} z^{-(n+l)} \left(S_{l} \otimes \bbbone\right)  
+\, \varepsilon_{1} \left(\bbbone\otimes S_{-n}\right)\,,
\ee
where $n\geq h+\alpha$. We can assume (by using the ambiguity in
defining $\alpha$) that $h+\alpha\geq 2$. 

The idea of the derivation is that we consider the quotient of the
fusion product of $\H$ with $\H_0$ by all states of the form 
\be
\label{quotient}
\Delta_{z,0} ( \A_{-} ) ( \H_0 \otimes \H )_f \,, \nonumber
\ee
where $\A_{-}$ is the algebra generated by all negative
modes. (In the conventional approach to fusion in terms of $3$-point
functions, all such states vanish if there is a highest weight vector
at infinity.) Using (\ref{chir20}), it is clear that we can identify
this quotient space with a certain subspace of 
\be
\left(\H_0 \otimes \H \right)_f / \Delta_{z,0} (\A_{-}) \left( \H_0 \otimes
\H \right)_{f} \subset \left( \H_0 \otimes \H^{(0)} \right)\,, \nonumber
\ee
where $\H^{(0)}$ is the highest weight space of the representation
$\H$. The idea is now to analyse this quotient space for the universal
highest weight representation $\H=\H_{univ}$ corresponding to the
twist $\alpha$, {\it i.e.} to use no property of
$\psi\in\H^{(0)}_{univ}$, other than that it is a highest weight
state. We can then identify this quotient space with a certain
quotient of the vacuum representation $\H_0$, thus defining $A(\H_0)$,
\be
\left( A(\H_0) \otimes \H^{(0)}_{univ} \right) = 
\left(\H_0 \otimes \H_{univ} \right)_f / 
\Delta_{z,0} (\A_{-}) \left( \H_0 \otimes \H_{univ} \right)_f  \,.
\ee
We should mention that $A(\H_0)$ will depend on the set of twists
$\alpha$ for each of the holomorphic fields $S$. 

The crucial ingredient we shall be using is the observation that,
because of (\ref{tilde}) and (\ref{cotrans}), 
$$\widetilde{\Delta}_{0,-z}(\tS_{-m})
\left( \H_0 \otimes\H \right)_{f}
$$
is for $m\geq h+\alpha$ in the space by which we quotient
(\ref{quotient}). In more detail we have
\begin{eqnarray}
{\displaystyle \widetilde{\Delta}_{0,-z}(\tS_{-m})} & = &
{\displaystyle \sum_{l=-m+\alpha}^{\infty} 
\left( \begin{array}{c} -\alpha \\ l+m-\alpha \end{array} \right) 
z^{-(m+l)} \left(S_{l} \otimes \bbbone\right)  }
\nonumber \\
\label{chir30}
& &  {\displaystyle 
+\, \varepsilon_{1} \sum_{k=1-h-\alpha}^{\infty} 
\left( \begin{array}{c} m+k-1 \\ m-h-\alpha \end{array} \right)
(-1)^{k+h+\alpha-1} (-z)^{-(m+k)} \left(\bbbone\otimes S_{k}\right)\,,} 
\end{eqnarray}
where $m\geq h+\alpha$.
\smallskip

The structure of the space $A(\H_0)$ depends on whether
$\alpha+h\in\Zop$ or not. In the first case, the field $S$ has a zero
mode for the representations with twist $\alpha$, and we have to find
a formula for
\be
(\bbbone\otimes S_0) (\H_0 \otimes \psi) \hspace*{1cm} 
\mbox{mod} \;\;\;\; \Delta_{z,0} (\A_{-}) (\H_0 \otimes \H_{univ}) \,,
\nonumber 
\ee
in terms of modes acting on the left-hand factor in the tensor
product. To do this, we use the explicit formula (\ref{chir30}) with
$m=h+\alpha$, and then (\ref{chir20}) for $n=1-h-\alpha, \ldots , -1$
to rewrite terms of the form $(\bbbone\otimes S_{n})$. After a short
calculation we obtain
\be
\label{zero}
\varepsilon_1 (\bbbone\otimes S_0) \cong
\sum_{m=0}^{h-1} 
\left( \begin{array}{c} h-1 \\ m \end{array} \right)
z^{h-m} (S_{m-h} \otimes \bbbone) \,.
\ee
This is independent of $\alpha$ (as it should be), and reproduces
precisely the formula of \cite{EG}. Using this expression, the action
of the zero mode on the fusion product then becomes
\be
\Delta_{z,0}(S_0) \cong \Delta_{z,0}(\tS_0) = \sum_{m=0}^{\infty} 
z^{h-m} \left( \begin{array}{c} h \\ m \end{array} \right)
(S_{m-h}\otimes\bbbone) \,,
\ee
which generalises Zhu's product formula \cite{Zhu}. Here we have used
that $S_0$ differs from $\tS_0$ only by negative modes which 
follows from (\ref{tilde}).   

Next, in order to obtain the relations which characterise the quotient
space $A(\H_0)$, we repeat the above calculation for $m=h+\alpha+p$,
where $p\geq 1$, and also use (\ref{zero}) to replace the term 
of the form $(\bbbone\otimes S_0)$. We then find that, modulo terms in
the quotient, we have 
\begin{eqnarray}
0 & \cong & 
{\displaystyle \sum_{m=-h-p}^{-h} 
\left( \begin{array}{c} -\alpha \\ m+h+p \end{array} \right)
z^{-(h+\alpha+p+m)} (S_{m} \otimes \bbbone)} \nonumber \\
&  &
\label{relation}
{\displaystyle +
\left( \begin{array}{c} h+\alpha+p-1 \\ p \end{array} \right)
(-1)^{p+1} z ^{-\alpha + p}  (S_{-h} \otimes \bbbone)} \\
&  &
{\displaystyle + \sum_{m=1-h}^{\infty} 
\left[ C_{m,p} + 
\left( \begin{array}{c} h-1 \\ m+h \end{array} \right)
\left( \begin{array}{c} h+\alpha+p-1 \\ p \end{array} \right)
(-1)^{p+1} \right]
z^{-(h+\alpha+p+m)} (S_{m} \otimes \bbbone)\,,} \nonumber
\end{eqnarray}
where $C_{m,p}$ is given as
\be
C_{m,p} = 
\left( \begin{array}{c} -\alpha \\ m+h+p \end{array} \right)
+ \sum_{l=1-h-\alpha}^{M_{h,\alpha}} 
\left( \begin{array}{c} h+\alpha+p+l-1 \\ p \end{array} \right)
(-1)^{p}
\left( \begin{array}{c} l+h-1 \\ m+h-1 \end{array} \right)\,,
\ee
and $M_{h,\alpha}<0$ is the largest number of the form $r-h-\alpha$,
where $r\in\Zop$. (In the present case, $M_{h,\alpha}=-1$.)
In general, this formula is rather complicated. However, for $p=1$,
using the same identities as in \cite{EG}, the expression simplifies
to 
\be
0 \cong z^{-\alpha} \sum_{m=0}^{\infty}
\left( \begin{array}{c} h \\ m \end{array} \right)
z^{-m} (S_{m-1-h} \otimes \bbbone) \,,
\ee
which gives conditions independent of $\alpha$. This generalises  
the formula of Zhu for $O(\H_0)$ \cite{Zhu}. 
\smallskip

If $\alpha+h\not\in\Zop$, then the field $S$ has no zero mode (for this
twist). In this case there is no analogue of the formula (\ref{zero}), 
but we can similarly do the calculation which lead to (\ref{relation})
for $p\geq 0$. Then the result is
\begin{eqnarray}
0 & \cong & 
{\displaystyle \sum_{m=-h-p}^{-h} 
\left( \begin{array}{c} -\alpha \\ m+h+p \end{array} \right)
z^{-(h+\alpha+p+m)} (S_{m} \otimes \bbbone)} \nonumber \\
&  &
{\displaystyle + \sum_{m=1-h}^{\infty} 
C_{m,p} z^{-(h+\alpha+p+m)} (S_{m} \otimes \bbbone)\,,}
\end{eqnarray}
where $C_{m,p}$ is given as above. For $p=0$, the formula simplifies
to 
\be
0 \cong 
\sum_{m=0}^{\infty}
\left( \begin{array}{c} M_{h,\alpha} + h \\ m \end{array} \right)
z^{-(\alpha+m)} (S_{m-h} \otimes \bbbone) \,.
\ee
This generalises the formula of \cite{DLM} (which was only derived for
integer $h$). We note that $M_{h,\alpha} + h$ is independent of the
ambiguity in choosing $\alpha$, {\it i.e.} invariant under
$\alpha\mapsto\alpha + r$, where $r\in\Zop$.

\subsection{Special Subspaces}
\renewcommand{\theequation}{3.2.\arabic{equation}}

It was observed by Nahm \cite{Nahm} that for untwisted representations
\be
\label{ineq}
\left(\H_1 \otimes \H_2 \right)_f / \Delta_{z,0} (\A_{-}) \left( \H_1
\otimes \H_2 \right)_{f} \subset \left( \H_1^s \otimes \H_2^{(0)}
\right)\,, 
\ee
where $\H_1^s$ is the special subspace of $\H_1$,
and $\H_2^{(0)}$ is the space of highest weight vectors in $\H_2$. 
The special subspace can be defined as the quotient space 
\be
\label{special}
\H_1^s = \H_1  / \A_{--} \H_1 \,,
\ee
where $\A_{--}$ is the algebra generated by the modes which do not
annihilate the vacuum, {\it i.e.} $\A_{--}$ is generated by
$S_{-n}, n\geq h$. Representations whose special subspace is finite
dimensional are called {\it quasirational}. It follows from
(\ref{ineq}) that the fusion product of a highest weight
representation with a quasirational representation contains only
finitely many highest weight representations.  

It follows from (\ref{chir10}, \ref{chir20}, \ref{chir30}) that
(\ref{ineq}) also holds if $\H_1$ is an untwisted representation, and
$\H_2$ is any twisted highest weight representation. It also makes
sense to extend the above definition (\ref{special}) and the
definition of quasirationality to cover the twisted case as well. In
particular, we can choose in (\ref{chir21}) $\alpha_i$ such that
$1\geq h+\alpha>0$ (so that $n=h+\alpha$ is the first negative mode)
and $1\geq h+\alpha_2>0$ (so that for $z_2=0$ and $n=h+\alpha$,
$\bbbone\otimes S_{-(h+\alpha_2)}$ is the only negative mode on the
right-hand-side). Then $1-h-\alpha_1 > -h$, and it follows by the same
argument as in Nahm \cite{Nahm} that the fusion product of any
quasirational representation with any highest weight representation
contains only finitely many subrepresentations, irrespective of
whether they are twisted or untwisted.

\section{Fusion in the $N=1$ algebra}
\renewcommand{\theequation}{4.\arabic{equation}}
\setcounter{equation}{0}

In this (and the following) chapter we want to use the explicit
formulae to (re)derive the fusion rules for the Neveu-Schwarz (NS) and
Ramond (R) sector of the $N=1$ and $N=2$ superconformal algebra. We
shall only consider the case of quasirational representations at
generic $c$. The fusion rules for the ``minimal'' models (at specific
values of the central charge) follow from these calculations if the
relevant representations are identified.

In this section we analyse the simpler case of the $N=1$ algebra. This 
algebra is generated by the Virasoro algebra $\{L_{n}\},\, n\in\bbbz$,
and the modes of the superfield $G$ with $h=\thalf$, subject to the
relations 
\begin{eqnarray}
{}[ L_{m}, L_{n} ] & = & {\displaystyle (m-n) L_{m+n} 
+ \frac{1}{12}\, c\, n\, (n^{2} -1)\, \delta_{n,-m}} \nonumber \\
{}[ L_{n}, G_{r} ] & = & {\displaystyle
\left( \half n - r \right) G_{n+r}} \\
\{ G_{r}, G_{s} \} &= & {\displaystyle 2 L_{r+s} 
+ \frac{1}{3} c \left( r^{2} - \frac{1}{4} \right) \delta_{r, -s}\,.}
\end{eqnarray}
The untwisted (NS) sector has $r\in\Zop + \half$, whereas the twisted
(R) sector has $r\in\Zop$. 

Let us parametrise as in \cite{GMTW} 
\be
c(t) = {15 \over 2} - {3 \over t} - 3t \,.
\ee
The representations we are concerned with are highest weight
representations of the chiral algebra, {\it i.e.} representations
generated by the action of the algebra from a state which is
annihilated by the action of the positive modes. In the NS sector,
there is only one zero mode ($L_0$), and the representation is uniquely
determined by the eigenvalue of the highest weight vector with respect
to $L_0$. We usually denote this highest weight state by $|h\rangle$,
where $h$ is the eigenvalue of $L_0$. If $h$ is of the form
\be
h_{p,q}(t) = {(1 - pq) \over 4} + (q^2 - 1) {t \over 8} 
+ {(p^2 - 1) \over (8t)} \,,
\ee
where $p,q\in\Zop$ and $p+q \in 2 \Zop$, then the representation
generated from the corresponding state has a singular vector at level
$pq /2$. The case $p=q=1$ corresponds to the vacuum representation
$h=0$; the first non-trivial case is $(p,q)=(1,3)$ with singular
vector 
\be
\left(G_{-\thalf} - {1 \over t} L_{-1} G_{-\half} \right) 
|h_{1,3}(t) \rangle 
\ee
and $(p,q)=(3,1)$ with singular
vector 
\be
\left(G_{-\thalf} - t L_{-1} G_{-\half} \right) 
|h_{3,1}(t) \rangle \,.
\ee
To determine the special subspace of the representations corresponding
to $(1,3)$ and $(3,1)$ we use the singular vector equations to
eliminate $L_{-1} G_{-\half} |h\rangle$, and the $G_{-\half}$
descendant of the singular vector equation to eliminate
$L_{-1}L_{-1}|h\rangle$. It then follows that the dimension is 
in both cases three. 

We shall also consider below the NS representation $(2,2)$ which has a
singular vector at level two
$$
\left(L_{-1}^2 - {4 \over 3} h_{(2,2)}(t) L_{-2} - G_{-\thalf}
G_{-\half} \right) |h_{2,2}(t) \rangle \,,
$$
and whose special subspace is two dimensional.

In the Ramond sector, there are two zero modes, and the
highest weight state is uniquely characterised by
\be
G_0 |\lambda\rangle = \lambda |\lambda\rangle \,,
\ee
as the commutation relations then imply 
$L_0 |\lambda\rangle = (\lambda^2 + c/24) |\lambda\rangle$. If
$\lambda$ is of the form
\be
\lambda_{p,q}(t) = { p - qt \over 2 \sqrt{2t}} \,,
\ee
where $p,q\in\Zop$ and $q-p$ odd, then the corresponding
representation has a singular vector at level $pq/2$. The first most
simple cases are $(p,q)=(1,2)$ with singular vector
\be
\left(L_{-1} + \sqrt{\frac{t}{2}} G_{-1} \right)| \lambda_{1,2}(t)\rangle = 0
\ee
and $(p,q)=(2,1)$ with singular vector
\be
\left( L_{-1} - {1 \over \sqrt{2t}} G_{-1} \right) 
| \lambda_{2,1}(t)\rangle = 0 \,.
\ee
We should note that the representation corresponding to $\lambda$ and
$-\lambda$ define the same vertex operator in correlation functions:
it follows from (\ref{action}) that the short distance expansion of
$G$ with the highest weight state has a branch cut. If we move $G$
once around the highest weight state, all coefficients change sign,
and thus, in particular, the eigenvalue $\lambda$ itself. This point
was overlooked  in \cite{GMTW}.

To determine the special subspace for these two representations, we
use the singular vector equation to eliminate $G_{-1}|\lambda\rangle$,
and the $L_{-1}$ and $G_{-1}$ descendants of the singular vector
equation to eliminate $L_{-1} G_{-1}|\lambda\rangle$ and
$L_{-1}L_{-1}|\lambda\rangle$. This then demonstrates that the
dimension of the special subspace is two for both cases.
\medskip

To illustrate how we can calculate the fusion products, let us
consider the situation, where we have a NS field at
$z_1=z$, and a R field at $z_2=0$. Choosing $\alpha_1=0$,
$\alpha_2=\alpha=1/2$, the comultiplication formulae simplify to
\be
\Delta_{z,0}(\tG_{n}) = \sum_{l=-1/2}^{\infty}
\left( \begin{array}{c} n+1/2 \\ l+1/2 \end{array} \right)
z^{n-l} (G_{l}\otimes\bbbone) 
+ \varepsilon_1 (\bbbone\otimes G_{n})\,,
\ee
where $n\geq -1$, and
\be
\Delta_{z,0}(\tG_{-n}) =  \!\! \sum_{l=-1/2}^{\infty}
\left( \begin{array}{c} n+l-1 \\ l+ 1/2 \end{array} \right)
z^{-(n+l)}
(-1)^{l+1/2} (G_{l}\otimes\bbbone) 
+ \varepsilon_1 (\bbbone\otimes G_{-n})
\ee
for $n\geq 2$. 

Let us now consider the case, where the representation at $z$ is any
NS highest weight representation, and the R representation at $0$ is 
$(1,2)$ or $(2,1)$. The singular vector equation implies that
\begin{eqnarray}
\label{rel1}
0 & = &
{\displaystyle \langle \bbbone\otimes L_{-1} \rangle + 
\alpha \langle \bbbone\otimes G_{-1} \rangle } \nonumber \\
& = &
{\displaystyle - z^{-1} \kappa \langle \bbbone\otimes \bbbone \rangle + 
\alpha \langle \bbbone\otimes G_{-1} \rangle\,, }
\end{eqnarray}
where $\alpha=\sqrt{t/2}$ for $(p,q)=(1,2)$ and
$\alpha=-\sqrt{1/2t}$ for $(p,q)=(2,1)$. Here the modes act on the
highest weight states of the corresponding representations. In the
second line we have used the standard relation (see {\it e.g.}
\cite{MG1}) to replace the $L_{-1}$ mode by inserting $\Delta(L_0)$;
$\kappa$ is then 
\begin{equation}
\label{kappa}
\kappa = h_3 - h_1 - h_2 \,,
\end{equation}
where $h_3$ is the conformal weight of the state at infinity, $h_1$
the conformal weight of the state at $z$, and $h_2=h_{(1,2)}$ or 
$h_2=h_{(2,1)}$. We can also use the equation coming from the $G_{-1}$
descendant of the singular vector equation to give
\begin{eqnarray}
\label{rel2}
0 & = &
{\displaystyle \langle \bbbone\otimes G_{-1} L_{-1} \rangle + 
\alpha \langle \bbbone\otimes G_{-1}^2 \rangle } \nonumber \\
 & = &
{\displaystyle - \half \langle \bbbone\otimes G_{-2} \rangle 
+\langle \bbbone\otimes L_{-1} G_{-1} \rangle + 
\alpha \langle \bbbone\otimes L_{-2} \rangle } \nonumber \\
& = &
{\displaystyle - z^{-1} (\kappa-1) \langle \bbbone\otimes G_{-1} \rangle
- \half \langle \bbbone\otimes G_{-2} \rangle
+ \alpha \langle \bbbone\otimes L_{-2} \rangle \,.}
\end{eqnarray}
Using the comultiplication of $L_{-2}$, it is easy to see that 
\be
\label{L2}
\langle \bbbone\otimes L_{-2} \rangle = (h_1 - \kappa) z^{-2} 
\langle \bbbone\otimes \bbbone \rangle \,.
\ee
Furthermore, from 
the comultiplication of $G_{-2}$ and $G_{-1}$ we learn that
\begin{eqnarray}
0 & = &
{\displaystyle z^{-\thalf} \langle G_{-\half} \otimes\bbbone \rangle + 
\varepsilon_1 \langle \bbbone\otimes G_{-2} \rangle \,,} \nonumber \\
0 & = &
{\displaystyle z^{-\half} \langle G_{-\half} \otimes \bbbone\rangle 
+ \varepsilon_1 \langle \bbbone\otimes G_{-1} \rangle \,,}
\end{eqnarray}
and this implies
\be
\label{G2}
\langle \bbbone\otimes G_{-2} \rangle = 
z^{-1} \langle \bbbone\otimes G_{-1} \rangle \,.
\ee
Taking (\ref{L2}), (\ref{G2}) and (\ref{rel2}) together, we find then
that 
\be
\label{rel3}
0 = - z^{-1} \left( (\kappa -1) + \half \right) 
\langle \bbbone\otimes G_{-1} \rangle + \alpha (h_1 - \kappa) z^{-2}
\,.
\ee
Putting this together with (\ref{rel1}), we obtain
\be
z^{-2} \left[ \alpha^2 (h_1 - \kappa) - \kappa (\kappa - \half)
\right] = 0 \,.
\ee
Thus a necessary condition for the fusion to be allowed is that the
bracket $[\ldots ]$ vanishes. This gives rise to the
relations\footnote{As always, we can only derive necessary conditions
for the fusion rules in this way. To derive sufficient conditions, a
much more detailed analysis is necessary. We also assume that all
fusion products are completely reducible --- this is justified for
generic $c$.}
\begin{eqnarray}
\label{res1}
(p,q)_{NS} \otimes (1,2)_R & = & (p,q+1)_R \oplus (p,q-1)_R \nonumber\\
(p,q)_{NS} \otimes (2,1)_R & = & (p+1,q)_R \oplus (p-1,q)_R \,.
\end{eqnarray}
\medskip

Using similar techniques we can derive the restrictions for the
fusion with another Ramond field. The result is identical to
(\ref{res1}), if we replace the suffices $NS$ on the left-hand side
and $R$ on the right-hand-side by $R$ and $NS$, respectively. We can
also determine the restrictions coming from the fusion of the $(1,3)$,
$(3,1)$ and $(2,2)$ field in the NS sector. We obtain (after similar
calculations) 
\begin{eqnarray}
\label{res2}
(1,3)_{NS} \otimes (p,q)_U & = & (p,q)_U \oplus (p,q+2)_U 
\oplus (p,q-2)_U  \nonumber\\
(3,1)_{NS} \otimes (p,q)_U & = & (p,q)_U \oplus (p+2,q)_U 
\oplus (p-2,q)_U  \nonumber\\
(2,2)_{NS} \otimes (p,q)_U & = & (p+1,q+1)_U \oplus (p-1,q-1)_U 
\,,
\end{eqnarray}
where $U=NS$ or $U=R$. Taking all of these results together and using
the associativity and commutativity of the fusion product, we can
give a compact formula for all different fusion products
\be
(p_1,q_1) \otimes (p_2,q_2) = \bigoplus_{p=|p_1 - p_2| + 1}^{p=p_1+p_2-1}
\bigoplus_{q=|q_1 - q_2| + 1}^{q=q_1+q_2-1} (p,q) \,,
\ee
where $p$ and $q$ are integers, and the sum is over every other
integer. The representation denoted by $(p,q)$ is in the NS sector if
$p+q \in 2\Zop$, and in the R sector otherwise. This reproduces the
results of \cite{Eich,SS,GMTW}.

\section{Fusion in the $N=2$ algebra}
\renewcommand{\theequation}{5.\arabic{equation}}
\setcounter{equation}{0}

In this section we want to analyse the fusion of quasirational
representations of the $N=2$ superconformal algebra for generic $c$.
Let us recall that this algebra is generated by the Virasoro algebra
$\{L_{n}\}$, the modes of an $U(1)$-current with $h=1$ $\{T_{n}\}$ and
the modes of two superfields of conformal dimension $h=\thalf$,
$\{G^{\pm}_{r}\}$, subject to the relations \cite{BFK}
\begin{equation}
\begin{array}{ccl}
{\displaystyle
\left[ L_{m}, G^{\pm}_{\alpha}\right]} & = &
{\displaystyle
\left(\half m - \alpha \right)\; G^{\pm}_{\alpha+m} } \nonumber \\
\vspace{0.2cm}

{\displaystyle
\left[ L_{m}, T_{n} \right] } & = &
{\displaystyle
- n \; T_{m+n}} \nonumber \\
\vspace{0.2cm}

{\displaystyle
\left[ T_{m},T_{n} \right]} & = & 
{\displaystyle \ct\; m \;\delta_{m,-n}} \nonumber \\
\vspace{0.2cm}

{\displaystyle
\left[ T_{m}, G^{\pm}_{\alpha} \right] } & = &
{\displaystyle \pm G^{\pm}_{\alpha+m} } \nonumber \\
\vspace{0.2cm}

{\displaystyle
\left\{ G^{\pm}_{\alpha}, G^{\pm}_{\beta} \right\}} & = & 
0 \nonumber \\
{\displaystyle
\left\{ G^{+}_{\alpha}, G^{-}_{\beta} \right\}} & = &
{\displaystyle
2 \;  L_{\alpha + \beta} + (\alpha - \beta) \; T_{\alpha + \beta}
+ \ct \;\left( \alpha^{2} - \frac{1}{4} \right) \delta_{\alpha,-\beta}, }
\end{array}
\end{equation}
where $\ct = c/3$. In the NS and R sector, the modes of $L_m$ and
$T_n$ are both integral, whereas the modes of $G^{\pm}_r$ are
half-integral in the NS sector, and integral in the R sector. There
also exists the so-called twisted sector, where the modes of $T_n$ and
$G^2_n$ are half-integral, those of $G^{1}_r$ and $L_n$ are
integral, and where $G^{\pm}=G^1 \pm i G^2$ \cite{BFK}. Here 
we shall only discuss the NS and the R sector.

It was observed in \cite{SchS} that there exists a family of
automorphisms $\alpha_\eta: \A \rightarrow \A$ which map the chiral
algebra to itself. They are explicitly given as 
\begin{eqnarray}
\label{auto}
\alpha_\eta(G^{\pm}_r) & = & G^{\pm}_{r\mp \eta} \nonumber \\
\alpha_\eta(L_n) & = & L_n - \eta \; T_n + {1 \over 2} \;\eta^2 \;\ct \;
\delta_{n,0} \\
\alpha_\eta(T_n) & = & T_n - \eta \; \ct \; \delta_{n,0} \,. \nonumber
\end{eqnarray}
For $\eta\in\Zop$, this gives an automorphism of each of the sectors
of the algebra, whereas for $\eta=\Zop + \half$, it relates the NS and
the R sector to each other. We shall demonstrate in this section that
the fusion rules respect the automorphism symmetry (see
(\ref{autosym}) below). The fusion rules for the $(NS \otimes NS)$ and
the $(R\otimes R)$ case are in principle known \cite{MSS2} (although,
strictly speaking, only the unitary minimal case is discussed there),
but in order to see that the automorphism symmetry is respected
generically, certain subtleties have to be taken into account which
were not discussed in \cite{MSS2}. We shall also derive the fusion
rules for the $(NS \otimes R)$ case.
\smallskip

Following \cite{BFK}, we want to parametrise the central charge
\be
\ct = 1 - {2 \over m} \,.
\ee
We shall only discuss the situation, where $m$ is generic, 
{\it i.e.} not a positive integer $m\geq 2$. Then the theory is not
unitary, and thus, as has been shown in \cite{EG}, not rational. The
unitary fusion rules can be obtained from the ones discussed in this
paper by identifying certain representations (see for example
\cite{MSS2}). 

In the NS sector, there are two zero modes, and a highest weight
representation is uniquely determined by giving the eigenvalue $h$ and
$q$ of the highest weight vector with respect to the action of $L_0$
and $T_0$, respectively. We parametrise
\begin{eqnarray}
q_{NS}(j,k) & = & {\displaystyle {j -k \over m}} \nonumber \\
h_{NS}(j,k) & = & {\displaystyle {j k - {1\over 4} \over m} \,.} 
\end{eqnarray}
For $0<j, j\in\Zop+\half$ the corresponding highest weight
representation has a fermionic singular vector at level $j$, and
likewise for $k$. For $n:= m - (j+k) \in\Zop, n>0$, the highest weight
representation has a bosonic singular vector at level $n$
\cite{BFK,Doerr}. The case $j=k=\half$ is just the vacuum
representation. We shall only analyse representations which
have two independent null-vectors, and in particular, only the
representations for which at least one of $j$ and $k$ is a half-integer,
and the other one is either a half-integer or of the form $m-s$, where
$s$ is a half-integer. It will turn out that this set of representations
is closed under the automorphism and under fusion.

In the R sector, there are in addition the two fermionic zero modes 
$G^{\pm}_0$, and to specify a highest weight representation it is
necessary to give the action of these modes as well. We will only
consider highest weight representations whose highest weight space
is (at most) two-dimensional, and we denote the two states by 
$| h, q\pm \half\rangle$, where 
\be
\label{Rhigh}
\begin{array}{rclrcl}
G_0^+ | h, q + \half\rangle & = & 0\,, \hspace*{3.4cm} &
G_0^- | h, q + \half\rangle & = & 
\sqrt{2 h - \ct / 4}\;\; | h, q - \half\rangle \,, \\
G_0^+ | h, q - \half\rangle & = & 
\sqrt{2 h - \ct / 4}\;\;  | h, q + \half\rangle \,, &
G_0^- | h, q - \half\rangle & = & 0 \,.
\end{array}
\ee
We parametrise the eigenvalues as
\begin{equation}
\label{N2para}
\begin{array}{rcl}
q_{R}(j,k) & = & {\displaystyle {j -k \over m}} \vspace*{0.2cm} \\
h_{R}(j,k) & = & {\displaystyle {j k \over m}  
+ \frac{\ct}{8}\,.} 
\end{array}
\end{equation}
For $0\leq j$ and $j\in\Zop$, the corresponding highest weight
representation has a fermionic singular vector at level $j$, and
likewise for $k$. For $n:= m - (j+k) \in \Zop$ there is also a
(bosonic) singular vector at level $n$ \cite{BFK}. We observe that
for $jk =0$, the two highest weight states are not related by the
action of $G_0^{\pm}$, as the coefficients in (\ref{Rhigh}) vanish.  
In this case, we denote by $(j,k)_{\pm}$ the two different
representations generated from the highest weight states
corresponding to $q_{R}(j,k) \pm \half$. 
\smallskip

Before we start analysing the fusion rules, let us describe how the
automorphism (\ref{auto}) acts on the various representations. Let us
first consider the NS representation $(j,k)$, and consider the
representation defined by 
\be
\label{autorep}
\hat{a}  | (j,k) \rangle := \alpha_\eta(a) | (j,k) \rangle \,,
\ee
where $| (j,k) \rangle$ denotes an arbitrary state in the highest
weight representation corresponding to $(j,k)$, and $a$ is an element
in the chiral algebra. We denote this representation by 
$$
(\hat{j},\hat{k}) = \alpha_\eta (j,k) \,.
$$

Let us first consider the case $\eta=1$. Then evaluating
(\ref{autorep}) on the highest weight state of the representation
$(j,k)$, we find that
\begin{eqnarray}
\hat{G}^{+}_{\half} | h(j,k) \rangle & = & G^{+}_{-\half} | h(j,k)
\rangle \,, \nonumber \\
\hat{G}^{-}_{\half} | h(j,k) \rangle & = & G^{-}_{+\thalf} | h(j,k) 
\rangle =0 \,,
\end{eqnarray}
and thus that (generically) the highest weight state of the
representation $(\hat{j},\hat{k})$ is $G^{+}_{-\half} | h(j,k)
\rangle$, rather than $| h(j,k)\rangle$ itself. We can then evaluate
$\hat{L}_0$ and $\hat{T}_0$ on this state, and find that \cite{Od}
\be
\alpha_{1} (j,k)= (j+1, k-1) \,.
\ee

In the above discussion something special happens for $k=1/2$, as then
\be
G^{+}_{-\half} \left| h\left(j,\half\right)\right\rangle = 0 \,.
\ee
Thus the state $| h(j,k)\rangle$ is the highest weight state
itself, and we find 
\be
\alpha_{1} \left(j,\half\right)= \left(\half, m-j-1\right) \,.
\ee
The same phenomena occurs for $\eta=-1$,
\be
\alpha_{-1} (j,k) = \left\{
\begin{array}{ll}
(j-1, k+1) & \mbox{if $j\neq \half$\,,} \\
(m-k-1,\half) & \mbox{if $j=\half$.}
\end{array}
\right. \ee

In the Ramond sector, similar considerations lead to
\be 
\alpha_{1} (j,k)_{+} = (j+1, k-1)_{-} \,,
\ee
and
\be
\alpha_{1} (j,k)_{-} = \left\{
\begin{array}{ll}
(j+1, k-1)_{-} & \mbox{if $j k\neq 0$\,,} \\
(1,m-j-1) & \mbox{if $j k =0$\,,}
\end{array}
\right. \ee
and similarly,
\be
\alpha_{-1} (j,k)_{-} = (j-1, k+1)_{+} \,,
\ee
and 
\be
\alpha_{-1} (j,k)_{+} = \left\{
\begin{array}{ll}
(j-1, k+1)_{+} & \mbox{if $j k \neq 0$\,,} \\
(m-k-1,1) & \mbox{if $j k =0$.}
\end{array}
\right. \ee
As a consistency check, we note that 
$$\alpha_{-1} \alpha_1(j,0)_{-}
= \alpha_{-1} (1,m-j-1) = (0,m-j)_{+}=(j,0)_{-}\,. $$

Finally, for $\eta=\pm \half$, we have maps from the NS sector in the
R sector,
\begin{eqnarray}
\alpha_{\half} (j,k) & = & \left(j+\half, k-\half\right)_{-} \,,\nonumber \\
\alpha_{-\half} (j,k) & = & \left(j-\half, k+\half\right)_{+} \,,\nonumber
\end{eqnarray}
and, conversely, maps from the R sector into the NS sector
\be
\alpha_{\half} (j,k)_{+} = \left(j+\half, k-\half\right) \,, 
\hspace*{1cm}
\alpha_{\half} (j,k)_{-} = \left\{
\begin{array}{ll}
\left(j+\half, k-\half\right) & \mbox{if $j k \neq 0$\,,} \\
\left(\half, m-j+k-\half\right) & \mbox{if $j k = 0$\,,}
\end{array} \right.
\ee
\be
\alpha_{-\half} (j,k)_{-} = \left(j-\half, k+\half\right) \,,
\hspace*{1cm} 
\alpha_{-\half} (j,k)_{+} = \left\{
\begin{array}{ll}
\left(j-\half, k+\half\right) & \mbox{if $j k \neq 0$\,,} \\
\left(m-k+j-\half,\half\right) & \mbox{if $j k = 0$\,.}
\end{array} \right.
\ee
\medskip

Let us now turn to analysing the fusion rules. First of all, from the
analysis of the $u(1)$ subalgebra, it follows that the eigenvalue of
the $T_0$-operator is additive under fusion. However, it is not clear
which states in the tensor product give a contribution with a highest
weight space at infinity, and therefore the quantum numbers
corresponding to the highest weight states do not always add up. In
fact, in general there are three different classes labelled by
$$
j_3 - k_3 = (j_1 - k_1) + (j_2 - k_2) + \delta m \,,
$$
where $\delta$ is either $0$, whence the fusion is called {\it even},
or $\delta = \pm 1$, whence the fusion is called {\it odd}
\cite{MSS1,MSS2}.  

The (even) fusion rules of a generic NS representation with the
$(\thalf,\half)$ and the $(\half,\thalf)$ representation have
been determined before using the comultiplication formula
\cite{MG2}. The result is
\be
\begin{array}{lcl}
\label{evenNS}
{\displaystyle
(j,k) \otimes \left(\thalf,\half\right)} & = & 
{\displaystyle (j+1,k) \oplus (j,k-1)\,,} \nonumber 
\vspace*{0.3cm}\\
{\displaystyle (j,k) \otimes \left(\half,\thalf\right)} & = & 
{\displaystyle (j,k+1) \oplus (j-1,k)} \,.
\end{array}
\ee
For generic $(j,k)$, this gives already the whole fusion rule, as the
odd fusion rule is forbidden. (This can be seen by combining the
restrictions coming from the two independent singular vectors. We
shall describe this is in detail in appendix~A.) The only values for
which an odd fusion is allowed is $k=\half$ in the first line, and
$j=\half$ in the second. For these values of $j$ (or $k$), the
representation $(j,k)$ has a singular vector at level $\half$, and
this will rule out one of the two representations in the fusion
product (\ref{evenNS}). (As a matter of fact, the representation which
would contain a negative value for $j_3$ or $k_3$ is absent.) Taking
this together, the fusion rules are (\ref{evenNS}), if $k\neq \half$,
or $j\neq \half$, respectively, and
\be
\begin{array}{lcl}
\label{31NS}
{\displaystyle \left(j,\half\right) \otimes 
\left(\thalf,\half\right)} & = & 
{\displaystyle \left(j+1,\half\right) \oplus \left(\half,m-j\right)\,,} 
\vspace*{0.3cm} \\ 
{\displaystyle \left(\half,k\right) \otimes \left(\half,\thalf\right)} 
& = & 
{\displaystyle \left(\half,k+1\right) \oplus \left(m-k,\half\right)} \,.
\end{array}
\ee
Furthermore, if both $j=k=\half$, then only the first term on the
right hand side survives. (This is consistent with $(\half,\half)$
being the vacuum representation.) 

Using the commutativity and associativity of the fusion product, 
these results are already sufficient to derive restrictions for 
the general fusion rules. We find
\be
\label{N2gen}
(j_1,k_1) \otimes (j_2,k_2) = 
\bigoplus_{j=\max(j_2-k_1,j_1-k_2) + \half}^{j_1+j_2-\half}
\left[\Bigl(j,j-(j_1+j_2) + (k_1+k_2)\Bigr)\right] \,,
\ee
where $j_l$ and $k_m$ are positive half-integers, and
\be
[(j,k)] = \left\{
\begin{array}{cl}
(j,k) & \mbox{if $j,k>0$\,,} \\
(-k,m-j) & \mbox{if $j>0, k<0$\,,} \\
(m-k,-j) & \mbox{if $k>0, j<0$\,.}
\end{array}
\right. 
\ee

Similarly, we can determine the fusion rules involving the field
$(\half,m-\fhalf)$
\be
\label{1mNS}
(j,k) \otimes \left(\half,m-\fhalf\right) = \left\{
\begin{array}{ll}
(j+2,k-1) \oplus (j+1,k-2) & \mbox{for $k \neq \half,\thalf$\,,}
\vspace*{0.2cm} \\
(j+2,k-1) \oplus (2-k,m-j-1) & \mbox{for $k =\thalf$\,,}
\vspace*{0.2cm} \\
(1-k,m-j-2) \oplus (2-k,m-j-1) & \mbox{for $k =\half$\,,} 
\end{array}
\right. \ee
and
\be
\label{m1NS}
(j,k) \otimes \left(m-\fhalf,\half\right) = \left\{
\begin{array}{ll}
(j-1,k+2) \oplus (j-2,k+1) & \mbox{for $j \neq \half,\thalf$\,,}
\vspace*{0.2cm} \\
(j-1,k+2) \oplus (m-k-1,2-j) & \mbox{for $j =\thalf$\,,}
\vspace*{0.2cm}  \\
(m-k-2,1-j) \oplus (m-k-1,2-j) & \mbox{for $j =\half$\,.} 
\end{array}
\right. \ee
Again, further cancellations arise if $j=\half$ in (\ref{1mNS}) and
$k=\half$ in (\ref{m1NS}). 

It is not difficult to see that (\ref{31NS}) and(\ref{N2gen}) 
are covariant under the automorphism (\ref{auto}), 
{\it i.e.} 
\be
\label{autosym}
\alpha_{\eta_1} (j_1,k_1) \otimes 
\alpha_{\eta_2} (j_2,k_2) = 
\alpha_{\eta_1+\eta_2} \left((j_1,k_1)\otimes (j_2,k_2)\right)\,.
\ee
Indeed, for example applying the automorphisms with $\eta_1=0$ and
$\eta_2=-1$ to the first line of (\ref{31NS}), we get 
for $j,k\neq \half$, 
\begin{eqnarray}
(j,k) \otimes \alpha_{-1} \left(\thalf,\half\right) & = &
(j,k) \otimes \left(\half,\thalf\right) \nonumber  \\
& = & (j,k+1) \oplus (j-1,k) \nonumber  \\
& = & \alpha_{-1} \left((j+1,k) \oplus (j,k-1)\right) \nonumber \\
& = & \alpha_{-1} \left((j,k) \otimes \left(\thalf,\half\right) 
\right) \,,
\end{eqnarray}
and for $j=\half$, $k\neq \half$
\pagebreak

\begin{eqnarray}
\left(\half,k\right) \otimes \alpha_{-1} \left(\thalf,\half\right) & = &
\left(\half,k\right) \otimes \left(\half,\thalf\right) \nonumber \\
& = & \left(\half,k+1\right) \oplus \left(m-k,\half\right) \nonumber \\
& = & \alpha_{-1} \left(\left(\thalf,k\right) \oplus 
\left(\half,k-1\right)\right) \nonumber \\
& = & \alpha_{-1} \left(\left(\half,k\right) \otimes 
\left(\thalf,\half\right) \right) \,.
\end{eqnarray}
The other cases are similar. As these fusion rules already determine
the general case, this implies that (\ref{autosym}) holds in
general. We can therefore deduce the general fusion rules of
fields $(j,k)$, where $j$ and $k$ are either positive half-integers
or of the form $m-s$, where $s$ is a positive half-integer, from
(\ref{N2gen}) using the automorphism (\ref{autosym}). For example
we have for $k>j$
$$
\left(j,m-k\right) = 
\alpha_{j+\half} \left(k-j-\half,\half\right)\,,
$$
and so 
\be
(j_1,k_1) \otimes (j_2,m-k_2) = 
\bigoplus_{j=\max(k_2-k_1,j_1+j_2) + \half}^{j_1+k_2-\half}
\left[\left(j,j-(j_1+j_2) + (k_1-k_2) \right)\right] \,.
\ee
The other cases are analogous. All of this holds for generic $m$,
{\it i.e.} generic $c$. For integer $m$, representations of the
different types can be identified, and the fusion rules are restricted
by the intersection of the various separate rules.
\medskip

We also determined the fusion rules of the $NS\otimes R$ sector, and
the $R\otimes R$ sector; the explicit results of our calculations can
be found in appendix~B. The main feature of these results is that all
fusion rules are covariant under the automorphism, where now also
$\eta\in \half\Zop$ is admissible. As they are sufficient to derive
the general restrictions, it follows that the automorphism is
respected in general. The most general fusion rules are therefore
already described by (\ref{N2gen}), provided we use the automorphism
(\ref{autosym}) as described before.

\section*{Appendix}

\appendix

\section{$N=2$ calculations}
\renewcommand{\theequation}{A.\arabic{equation}}
\setcounter{equation}{0}

In this appendix we shall give some details for the derivation of the
simplest interesting case where a cancellation appears, the fusion of 
a NS field with $(\thalf,\half)$. 

The derivation of the even fusion rules was explained in
\cite{MG2}.\footnote{We should mention that the notation here 
differs from that in \cite{MG2} by $j \leftrightarrow k$.} For the odd
case, because of the null-vector
\be
\label{null1}
\Nu_1 = G^{+}_{-\half} \phi_{\thalf,\half} = 0 \,,
\ee
only one of the two possible odd fusion rules is non-trivial, the one
where $q_3=q_1+q_2-1$. Furthermore, the representation $(\thalf,\half)$
has a second null vector,
$$
\Nu_2 = \O_2 \phi_{\thalf,\half} =
\left( -\frac{1}{m}  G^{-}_{-\thalf} + L_{-1} G^{-}_{-\half}
+ T_{-1} G^{-}_{-\half} \right) \phi_{\thalf,\half} = 0 \,,
$$
and this implies the following restriction
\begin{eqnarray}
0 & = & \langle (\bbbone \otimes \O_2 ) \rangle \nonumber \\
& = & {\displaystyle 
z^{-1} \langle (\bbbone \otimes G^{-}_{-\half} ) \rangle
\left[ -\frac{1}{m} - \left(\kappa - \half\right) - q_1 \right]\,,}
\nonumber 
\end{eqnarray}
where $\kappa$ is as before (\ref{kappa}), and $q_1$ denotes
the $T_0$ eigenvalue of the highest weight state at $z=0$. (To derive
this equation, we have, as usual, considered the action of the
comultiplication $\Delta_{z,0}$ of $G^{-}_{-\thalf}, G^{-}_{-\half},
L_{-1}$ and $T_{-1}$ on the product, which vanishes in correlation
functions because of the highest weight property of the state at
infinity.) If the (odd) fusion is allowed, the correlation function  
does not vanish, and we must have that the bracket is zero.
This implies that
\begin{equation}
\label{kappares}
\kappa = \half - \frac{1}{m} - q_1 \,.
\end{equation}

Next we consider the condition coming from the 
$G^{-}_{-\half}\, G^{+}_{-\half}$ descendant of $\Nu_2$, which
simplifies to
\begin{eqnarray}
\Nu_3 & = & {\displaystyle \left( 
- \left( \frac{2}{m} + 2\right) \left( G^{-}_{-\fhalf} 
+ L_2 G^{-}_{-\half}\right) 
- \left( \frac{1}{m} -1 \right) 
T_{-2} G^{-}_{-\half} \right.} \nonumber \\
& & \;\;\;\; {\displaystyle \left.
+ 2 L_{-1}^2 G^{-}_{-\half}
+ 2 T_{-1} L_{-1} G^{-}_{-\half}
+ 2 L_{-1} G^{-}_{-\thalf} \right) \phi_{\thalf,\half} = 0 \,.}
\end{eqnarray}
Using the same techniques as before, this implies the equation
\begin{eqnarray}
\label{kapparel}
0 & = & {\displaystyle 
z^{-2} \langle (\bbbone \otimes G^{-}_{-\half} ) \rangle
\left[ \left( \frac{2}{m} + 2 \right) 
\left(1 + h_1 - \left(\kappa - \half \right) \right)
-  \left(\frac{1}{m} -1\right) q_1 \right. } \nonumber \\
& & \;\;\;\; {\displaystyle \left. 
- 2 \left(\kappa - \thalf\right) \left(\kappa - \half \right)
- 2 q_1 \left(\kappa - \half\right) - 
2 \left(\kappa - \thalf\right) \right]\,.}
\end{eqnarray}
We can now evaluate (\ref{kapparel}), plugging in the value for
$\kappa$ from (\ref{kappares}), and the equation then becomes
\begin{equation}
0 = \frac{ (1 + 2 j) (1 - 2 k) (1 + m)}{2 m^2} \,,
\end{equation}
where $(j,k)$ labels the $L_0$ and $T_0$ eigenvalues of the highest
weight state at $z$ as in (\ref{N2para}). We conclude that the odd
fusion is only allowed for $k=\half$, as we can restrict $j$ and $k$
to be positive half-integers or of the form $m-p$, where $p$ is a
half-integer. (As we mentioned before, we are considering here the
generic case, where $m$ is not an integer.)  For $k=\half$, we can
evaluate (\ref{kappares}), and this gives the fusion, as described in
(\ref{31NS}).
\smallskip

To see that only one of the even fusions is allowed for $k=\half$,
we observe that in this case, the field at $z$ has the null-vector
(\ref{null1}), and we thus have
\begin{eqnarray}
0 & = & {\displaystyle 
\langle \Delta_{z,0}(G^{+}_{-\half}) 
(\phi_{j,\half} \otimes G^{-}_{-\half} \phi_{\thalf,\half})
\rangle} \nonumber \\
& = & {\displaystyle 
2 \langle 
(\phi_{j,\half} \otimes L_{-1} \phi_{\thalf,\half}) \rangle}
\nonumber \\
& = & {\displaystyle
- 2 \kappa z^{-1} \langle 
(\phi_{j,\half} \otimes \phi_{\thalf,\half}) \rangle\,.} \nonumber
\end{eqnarray}
This implies for the even fusion rules
\begin{eqnarray} 
j_3 - k_3 & = & j +\half \nonumber \\
j_3 k_3 & = & \half j + \half \,, \nonumber
\end{eqnarray}
which has the unique solution $j_3 =j+1, k_3=\half$.

\section{Some $N=2$ fusion rules}
\renewcommand{\theequation}{B.\arabic{equation}}
\setcounter{equation}{0}

The results for the $(R\otimes NS)$ fusion are 
\be
\label{31NSR}
(j,k)_{-} \otimes \left(\thalf,\half\right) = \left\{
\begin{array}{ll}
(j+1,k) \oplus (j,k-1)_{-} & \mbox{if $j k\neq 0$\,,} \vspace*{0.2cm} \\
(j+1,k)_{-} \oplus (1,m+k-j) & \mbox{if $j k=0$\,,} 
\end{array} \right. 
\ee
\be
\label{1mNSR}
(j,k)_{-} \otimes \left(\half,m-\fhalf\right) = \left\{
\begin{array}{ll}
(j+2,k-1) \oplus (j+1,k-2)_{-} & \mbox{if $k\neq 0,1,\, j\neq 0$\,,}
\vspace*{0.2cm} \\
(j+2,k-1)_{-} \oplus (1,m+k-j-2) & \mbox{if $k=1,\, j\neq 0$ \,,}
\vspace*{0.2cm} \\
(1,m+k-j-2) \oplus (2,m+k-j-1) & \mbox{if $j k=0$\,,} 
\end{array} \right. 
\ee
\be
\label{11NSR}
(1,1) \otimes (j,k) = \left\{
\begin{array}{ll}
\left(j+\half,k+\half\right) \oplus 
\left(j-\half,k-\half\right)_{-} & \mbox{if $j\neq \half$\,,} 
\vspace*{0.2cm} \\
\left(j+\half,k+\half\right) \oplus \left(m-k+\half,0\right)_{-} 
& \mbox{if $j= \half$\,,}
\end{array} \right.
\ee
and
\be
\label{20NSR}
(2,0)_{-} \otimes (j,k) = \left\{
\begin{array}{ll}
(j+\thalf,k-\half) \oplus (j+\half,k-\thalf)_{-} 
& \mbox{if $k\neq \half$\,,} \vspace*{0.2cm} \\
(j+\thalf,k-\half)_{-} \oplus (1,m+k-j-1) & \mbox{if $k=\half$\,,}  
\end{array} \right. 
\ee
where for the R representation $(j,k)$, for which one of the labels
might be zero, a suffix $\pm$ has been included in order to indicate
which of the two representations is referred to. The fusion rules
of the fields $(\half,\thalf)$ and $(m-\fhalf,\half)$ can be 
obtained from (\ref{31NSR}) and (\ref{1mNSR}), using the obvious
symmetry.   

We also determined the following $(R \otimes R)$ fusion rules
\be
\label{11RR}
(1,1) \otimes (j,k)_{+} = \left\{
\begin{array}{ll}
(j+\half,k+\half) \oplus (j-\half,k-\half) & \mbox{if $j k \neq 0$\,,}
\nonumber \vspace*{0.2cm} \\
(j+\half,k+\half) \oplus (m-k+j+\half,\half) & \mbox{if $j k =0$ \,,}
\end{array} \right.
\ee
and
\be
\label{20RR}
(2,0)_{-} \otimes (j,k)_{-} = \left\{
\begin{array}{ll}
(j+\thalf,k-\half) \oplus (j+\half,k-\thalf) 
& \mbox{if $k\neq 1,0, \, j\neq 0$\,,} \vspace*{0.2cm} \\
(j+\thalf,k-\half) \oplus (\half,m+k-j-\thalf)
& \mbox{if $k=1, \, j\neq 0$\,,}  \vspace*{0.2cm} \\
(\thalf,m+k-j-\half) \oplus (\half,m+k-j-\thalf) 
& \mbox{if $j k=0$\,.}
\end{array} \right. 
\ee

{\bf Acknowledgements}

I would like to thank Adrian Kent for suggesting this problem and for
useful discussions. I acknowledge useful conversations with Matthias
D\"orrzapf, Wolfgang Eholzer, Peter Goddard and G\'erard Watts.  

This work was supported by a Research Fellowship of Jesus College,
Cambridge, and partly by PPARC and EPSRC, grant GR/J73322.

\end{document}